\documentclass[conference]{IEEEtran}
\IEEEoverridecommandlockouts
\pdfoutput=1
\usepackage{cite}
\usepackage{amsmath,amssymb,amsfonts}
\usepackage{algorithmic}
\usepackage{graphicx}
\usepackage{textcomp}
\usepackage{xcolor}
\usepackage{nicematrix}
\usepackage{blkarray}
\def\BibTeX{{\rm B\kern-.05em{\sc i\kern-.025em b}\kern-.08em
    T\kern-.1667em\lower.7ex\hbox{E}\kern-.125emX}}
\begin{document}

\title{Molecular Representations for AI in Chemistry and Materials Science: An NLP Perspective}
% A perspective of NLP Engineer towards Material Science Research
%Exploring  Material science using NLP techniques

\author{\IEEEauthorblockN{Sanjanasri JP, Pratiti Bhadra, N. Sukumar, Soman KP}
\IEEEauthorblockA{\textit{Center for Computational Engineering and Networking (CEN)} \\
\textit{Amrita Vishwa Vidyapeetham}\\
Coimbatore, India \\
jp\_sanjanasri@cb.amrita.edu}
% \and
% \IEEEauthorblockN{Pratiti Bhadra}
% \IEEEauthorblockA{\textit{Center for Computational Engineering and Networking (CEN)} \\
% \textit{Amrita School of Engineering}\\
% \textit{Amrita Vishwa Vidyapeetham}\\
% Coimbatore, India \\
% b_prartiti@cb.amrita.edu}
% \and
% \and
% \IEEEauthorblockN{Soman KP}
% \IEEEauthorblockA{\textit{Center for Computational Engineering and Networking (CEN)} \\
% \textit{Amrita School of Engineering}\\
% \textit{Amrita Vishwa Vidyapeetham}\\
% Coimbatore, India \\
% kp\_somani@amrita.edu}

}

\maketitle

\begin{abstract}

Deep learning, a subfield of machine learning, has gained importance in various application areas in recent years. Its growing popularity has led it to enter the natural sciences as well. This has created the need for molecular representations that are both machine-readable and understandable to scientists from different fields. Over the years, many chemical molecular representations have been constructed, and new ones continue to be developed as computer technology advances and knowledge of molecular complexity increases. This paper presents some of the most popular digital molecular representations inspired by natural language processing (NLP) and used in chemical informatics.
In addition, the paper discusses some notable AI-based applications that use these representations. This paper aims to provide a guide to structural representations that are important for the application of AI in chemistry and materials science from the perspective of an NLP researcher. This review is a reference tool for researchers with little experience working with chemical representations who wish to work on projects at the interface of these fields.

\end{abstract}


\section{Introduction}
Artificial Intelligence (AI) and Machine Learning have revolutionized various disciplines in the last few years, and cheminformatics is no exception~\cite{David:Molecular}. Cheminformatics is a scientific field that uses computational techniques, such as machine learning, to tackle various chemistry-related issues. These issues may involve predicting the properties of molecules, finding new ways for reactions to occur, synthesizing chemicals, and understanding how biologically active compounds work and how they can be developed into pharmaceutical drugs.

The conventional development of new drugs/molecules is a complex and time-consuming procedure. One of the most important reasons is that the generation/representation of molecules strongly relies on external expert knowledge. In this case, expert knowledge may consist of chemical compounds/fragments that may be ``mixed and matched" to form a collection of prospective molecules~\cite{Daniel:review}. 
More importantly, they may not possess the desired properties - or may possess other undesirable properties. Hence, the resultant compounds are challenging to synthesize. As a result, research into highly effective smaller molecules and materials has been hampered.

The collection of all conceivable chemical compounds that may be generated, regardless of whether they exist in nature or have been manufactured by humans, is referred to as chemical space. It encompasses all elements and their various combinations, as well as the different structural arrangements and chemical properties that can result from these combinations~\cite {Joshi:molecules}. The chemical space is vast and huge, with estimates of possible small molecules alone ranging in the billions or even trillions.

Exploring such vast and complex chemical space using traditional methods has allowed only a small portion to be studied~\cite{Bagal:MolGPT}. However, utilizing cutting-edge technologies such as machine learning and deep learning can accelerate the exploration of a more significant amount of chemical space. Researchers have made significant strides in drug discovery using AI techniques~\cite{David:Molecular}. However, fundamental tasks such as predicting the properties of molecules, retrosynthesis, and chemical reaction optimization need to be addressed to understand new drug functionality.

%Only a small portion of the known potential chemical space has been studied using traditional approaches, given the breadth and complexity of chemical space. However, when aligned with state-of-art technologies, the expedition in exploring more significant portions of the chemical space can be accelerated. The researchers have made substantial progress toward drug discovery using AI techniques.  However, predicting the property of molecules, retrosynthesis, chemical reaction optimization, etc., are some other fundamental tasks to explore the functionality of new drugs. 

It is necessary to represent molecules in a three-dimensional structure in a machine-readable format for the accurate functioning of the AI models. This format should accurately reflect the structural properties of molecules. It is worth mentioning that many recent AI-based molecular representation techniques are based on Natural Language Processing concepts~\cite{Bagal:MolGPT}.

Natural language processing (NLP) is a computer science field that encompasses linguistics and Artificial Intelligence (AI)~\cite{Jurafsky:2009}. NLP deals with interactions between machines and the natural languages of humans. AI is an umbrella term for computers that can comprehend human intelligence in understanding text and spoken words. NLP aims to transform text information into structured data to enhance the data's usability and the quality of decisions based on that data. Representation of words is the first and necessary step in NLP to start with any Machine Learning applications such as translation, chatbots, automatic grammar checking, etc. Such representations, commonly called word embedding, map the word or phrase from the text to real-value vectors~\cite{sanjana:2021}. These vectors elucidate the text's hidden information (semantics); semantics is the word's meaning~\cite{sanjana:autosar}.

Analogous to words in NLP, atoms form the smallest unit in materials science. Materials are treated as languages, atoms as words. Correct representation of word order forms a meaningful sentence/text. Likewise, the correct sequence of atoms determines the appropriate material or molecule. Symbolic representations of molecules and materials differ from words and sentences. However, the word is a simple sequential structure whose representation is straightforward, whereas a molecule is a physical collection of atoms arranged in 3D space. Therefore, the molecular representation (1D or 2D) should include all information such as bonds, branch start and end points, number of additional bonds, etc~\cite{Krenn:Selfies}.

This paper describes a guide to molecular representation using NLP. Section two highlights the difficulties of representing materials in a machine-readable format. Section three explores the various methods currently used in AI for chemistry and materials science, with a focus on their applications. Section four provides examples of AI-powered applications that utilize the chemical representations discussed in the study. The paper concludes with section five.

\section{Challenges in representing Molecules}

There are several challenges in representing molecules in chemistry and biology:

\begin{itemize}
    \item \textbf{Complexity:} Molecules can have many atoms and bonds, making them complex and difficult to represent.
    \item \textbf{3D structure:} Molecules have a 3D structure that can influence their properties and behavior. Representing this 3D structure accurately can be challenging.
    \item \textbf{Multiple conformations:} Molecules can exist in multiple conformations, which are different 3D arrangements of the atoms. Representing all possible conformations of a molecule can be computationally intensive.
    \item \textbf{Atom labeling:} The atoms in a molecule must be appropriately labeled for the representation to be accurate. This can be challenging when multiple atoms have the same elemental composition.
    \item \textbf{Hydrogen representation:} Hydrogen atoms can be explicit, represented as a separate atom, or implicit, not defined but understood to be present. Deciding how to handle hydrogens can affect the accuracy of the representation.
    \item \textbf{Small differences can have significant effects:} Small changes to a molecule's structure can lead to substantial changes in its properties and behavior. Representing these minor differences accurately can be challenging.
\end{itemize}

Hence, any effective machine-readable molecular representation must consider the challenges above for a successful application for downstream tasks such as drug discovery, chemical synthesis, etc.

There are many ways to represent molecules; the most suitable representation depends on the specific application. Some common ways to represent molecules include:

\section{Representation of Molecules}
\begin{figure}[htbp]
\centerline{\includegraphics[scale=1.0]{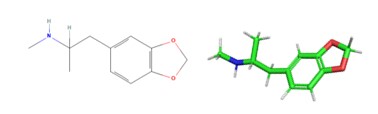}}
\caption{Two Dimensional and Three Dimensional Structure of 3,4-Methylenedioxymethamphetamine (MDMA) molecule }
\label{fig:2dpaper}
\end{figure}

Traditionally, molecules are represented on paper as a two-dimensional molecular structure, as shown in figure~\ref{fig:2dpaper} or as a chemical formula that represents a molecule using chemical symbols to represent the elements present and subscripts to indicate the number of atoms of each element.  For example, $C_{11}H_{15}NO_2$ is the molecular or chemical formula for 3,4-Methylenedioxymethamphetamine (MDMA). The chemical formula representation does not encode any other information related to the syntax and semantics of molecular structure.

In the machine learning era of cheminformatics, various methods have been developed to represent molecules in a data-driven way~\cite{Guo:Graph}. These include string-based, connection table, feature-based, and computer-learned representations. In this section, the most widely used molecular representation methods are discussed and grouped into two categories: a) String-Based, and b) Graph-Based.

\subsection{String-Based Representation}

In computer systems, the text is often represented using the American Standard Code for Information Interchange (ASCII) character encoding standard, which includes 128 characters such as letters, numbers, and symbols. These representations, called ``strings'', are more compact and more manageable for humans to read and write than other character encodings. Also, the string representation encourages the adaption of state-of-art AI techniques in NLP domain, such as embedding and transformers to represent molecules for computation~\cite{Guo:Graph}. Henceforth, molecules are represented as strings in ASCII format in natural science for computation.  The following subsection summarizes the widely used string representation. 

\subsubsection{SMILES}

A method for representing molecules as a single line of text is the Simplified Molecular Input Line Entry System (SMILES). The structural details of a molecule are encoded by SMILES using a set of rules. The guidelines follow the Left-Linear (LL1) Grammar standard in computer science~\cite{Weininger:SMILES}. To produce string derivation that may represent the molecular structure, this language can be thought of as a little computer program. The Backus-Naur form (BNF) is one of the most widely used standard notations for expressing computational grammar~\cite{Ullman:compiler}. Below is a grammatical example. 

\hfill \break
$<SMILES\_Line>         ::= <Atom> ( <Chain> | <Branch> )*$

$<Chain>          ::= ( <Dot> <Atom> | <Bond>? ( <Atom> | <Rnum>) )+$

$<Branch>         ::= ''('' ( ( <Bond> | <Dot> )? <Line> )+ ")"$

$<Atom>           ::=  <Organic\_symbol> | <Bracket\_atom>$

$<Rnum>   ::= <Digit> | "\%" <Digit> <Digit>$

\hfill \break 
A grammar expressed in BNF consists of three parts. A Non-Terminal is followed by an assignment operator $(::=)$, followed by one or more terminals and non-terminals. SMILES grammar has rules to transform each entity in molecular structure to grammar. For instance, branching in the molecular structure is defined within parentheses, while two matching numbers indicate ring closures; the number representing starting and closure of the ring. The SMILES representation for 3,4-Methylenedioxymethamphetamine (MDMA) molecule is \textit{``CC(CC1=CC2=C(C=C1)OCO2)NC''}.

Natural language processing techniques, including word embedding and recurrent neural networks, have been used to analyze SMILES representations as they consist of text~\cite{Weininger:SMILES}. The SMILES notation has been a popular tool in computational chemistry since 1980~\cite{Esben:SMILES}. Due to its extensive grammar rules, the SMILES notation is more challenging for complex molecular structures. With the advancement of deep learning in cheminformatics and the subsequent development of novel molecular string representations, the limitations of SMILES have also become more apparent.

The following are some significant issues with SMILES representation.

\begin{itemize}
    \item \textbf{Ambiguity:} Multiple molecules can have the exact SMILES representation or the same molecule can have multiple representations, leading to confusion. SMILES representation for MDMA molecule are \textit{``CC(CC1=CC2=C(C=C1)OCO2)NC'', ``CC(CC1CCC2C(C1)OCO2)NC'' and ``CNC(C)CC2CCC1OCOC1C2''}. This representation ambiguity is unsuitable while dealing with large-scale databases for any deep learning tasks. 
    
    Even the canonical form of SMILES which reduces the different representations of the same molecule to a unique, standard form, does not take R/S stereoisomerism into account, so each enantiomer of a compound will have the same canonical SMILES formula.
    
    \item \textbf{Limited expressiveness:} SMILES is a linear notation, which means it represents information about atoms and bonds only in a linear manner. This can, however, represent the spatial relationship between non-bonded atoms in 3D space and the isomeric form of molecules.  It should also be noted that a ring structure can have multiple ring-closure points. Rings with six bonds can each have a point of closure, such as those in six-membered rings. Therefore, a ring compound can be represented by a variety of SMILES strings that are equally valid. Since SMILES strings can be derived from any atom in a molecule, it is very common that many SMILES strings can represent the same structure, whether it has a ring or not. In this context, it is crucial to select a "unique SMILES" out of an array of potential ones for each molecule. CanonicalizationThis unique SMILES string, also referred to as 'canonical SMILES', accounts for stereochemistry but fails for unusual aromatic models.
    
    For example, the benzyl molecule can be expressed as 'c1ccccc1c2ccccc2' or 'c1ccccc1c1ccccc1'. In SMILES, a ring closure digit can be reused if desired, but this has the disadvantage of not representing the aromaticity property of the molecule.

    \item \textbf{Semantic Errors:} SMILES lacks a technique to guarantee that molecular strings adhere to physical and syntactic rules. The SMILE system can generate a string such as \textit{ CO (C} with an unpaired opening parenthesis that does not have a valid interpretation as a molecular graph. Such generated string (\textit{ CO (C})  is called Semantic error and involves strings that form valid graphs but do not reflect reasonable chemical structures.

    \item \textbf{Syntactic Errors:} A considerable portion of the generated SMILES strings have a high percentage of invalid compounds. They break fundamental chemical restrictions, such as the permissible number of valence bonds between atoms, or they are syntactically flawed, meaning they do not even conform to a molecular diagram. 
    For instance, the SMILE string (\textit{CO = C}) represents a molecular diagram with one oxygen atom bound by three bonds; this is an error since neutral oxygen can only make two bonds.
\end{itemize}

Despite these limitations, SMILES representations are still a workhorse in cheminformatics to study a smaller group of molecules.

\subsubsection{InChI}

International Chemical Identifier (InChI)~\cite{Heller:InChI} is another system for representing chemical compounds in a standardized, machine-readable format. The International Union of Pure and Applied Chemistry (IUPAC)~\cite{Leigh:IUPAC} developed this open-source molecular representation in 2013 to uniquely and persistently identify chemical substances across different databases and platforms. InChI can represent the structural information of various chemical compounds, including small molecules, polymers, and biomolecules. The InChI format is based on a layered approach, which allows for different levels of detail to be encoded depending on the desired use case. The standard InChI has a main layer that encodes the topology and stereochemistry of a molecule and an optional fixed-hydrogen layer that encodes the exact positions of all hydrogen atoms, unlike SMILES. InChI can also specify which hydrogen atoms are mobile and which are immobile. MDMA representation in the INCHI string is \textit{``1S/C11H15NO /c1-8(12-2)5-9-3-4-10-11(6-9)14-7-13-10/h3-4,6,8,12H,5,7H2,1-2H3"}.

Though InChI is widely used as a standard molecular representation format, it does have some limitations. Some of the drawbacks of InChI include:
\begin{itemize}
    \item \textbf{Length:} InChI can be quite long, especially for larger and more complex molecules. This can make it challenging to use in applications where space is limited.
    \item \textbf{Uniqueness:} While InChI is designed to provide a unique identifier for each chemical substance, there are some cases where multiple InChIs can be generated for the same molecule. This can lead to confusion and potential errors in databases and other applications.
    \item \textbf{Syntax Complexity:} InChI notation is not as simple as SMILES, making it harder for humans to read and interpret.

    \item \textit{Lack of explicit hydrogen representation:} The InChI notation does not explicitly represents hydrogen's position in the molecule, which can make it harder to use in applications that require that information.

    \item \textbf{Performance:} The InChI generation and parsing process can be computationally expensive, which can be an issue for large datasets or high-throughput applications.

\end{itemize}
    
To handle the  string length issue of InChI representation, the researchers have introduced a format called InChI Key. It is a short, unique, and fixed-length identifier derived from a molecule's InChI representation. It is designed to be a more compact and human-readable version of the InChI, which can be used as a primary key in databases and for searching and indexing purposes.

The InChI Key is generated by taking the first 14 characters of the main layer of the InChI and the 8th character of the InChI fixed-hydrogen layer. The key is a 27-character long, case-sensitive, alphanumeric code that represents the molecular structure of a compound and is unique for each chemical substance. The InChI key is highly useful in chemical databases, as it allows for quick identification of compounds and searching of chemical databases. InChI key for MDMA molecule is \textit{``SHXWCVYOXRDMCX-UHFFFAOYSA-N"}

Though InChI has proven to be a robust and versatile molecular representation format, it is not the best fit for all applications.

\subsubsection{DEEPSMILES}

Researchers have developed DeepSMILES to address some of the shortcomings of the SMILES notation, a commonly used format in computational chemistry. DeepSMILES is based on the SMILES notation but has additional features that allow for encoding more complex molecular information~\cite{Snchez:Inverse}.  

One of the main features of DeepSMILES is its ability to represent the 3D structure of a molecule, which is not possible with the original SMILES format. DeepSMILES uses a graph-based representation of a molecule, where nodes and bonds represent atoms are represented by edges. This allows for the encoding of the topology, stereochemistry, and even the explicit positions of hydrogen atoms. Additionally, DeepSMILES uses a deep learning-based algorithm to generate molecular representation, which efficiently handles large and complex molecular structures.

The DeepSMILES syntax avoids the problem of unbalanced parentheses by using only close parentheses, where the number of parentheses indicates the length of the branch. In addition, DeepSMILES avoids the problem of pairing ring closure symbols by using only a single symbol at the ring closure point, where the symbol indicates the ring size, i.e., the number of atoms that make up the ring. The DeepSMILES representation for the MDMA molecule is \textit{``CNCC)CC=CC=C(6)OCO5"}

DeepSMILES has been used in various applications such as drug design, Quantitative Structure-Activity Relationship (QSAR) modeling, and virtual screening. It resolves most cases of syntactical mistakes. However, it also has some limitations, including:

\begin{itemize}
    \item  \textbf{Limited applicability:} DeepSMILES is designed to handle 3D structures of molecules which is not a requirement in some specific fields such as natural product representation.

    \item \textbf{Lack of standardization} DeepSMILES is a relatively new format and it's not as widely adopted as other formats like SMILES and InChI. This can make it harder to find tools and resources that support it.

   \item \textbf{Size:} The size of DeepSMILES representation can be larger than other molecular representation formats like SMILES, which can be an issue for applications that require the storage of large datasets.
   
    \item  \textbf{Semantic Error:} Generate semantically incorrect strings, i.e., molecules that violate primary physical constraints.
\end{itemize}

The aforementioned factors point to a need for an even more robust molecular grammar~\cite{Tetko:transformers}. 

\subsubsection{SELFIES}

Simple Explicitly-Localized Formalism for Incredibly Easy Specification of Isomers and Elements or SELFIES is a molecular representation format that guarantees the chemical validity of the resulting molecule. SELFIES overcomes the drawbacks of all its predecessor in string representation. It considers information about branching, ring, and valence constraints and is designed to eliminate syntactically and semantically invalid molecules~\cite{Krenn:Selfies}.

SELFIES is based on an LL(1) formal grammar, with specific derivation rules and conventions to represent a molecule. It starts by defining the explicit position of each atom in the molecule and then encodes the connectivity between atoms and the properties of atoms and bonds. The SELFIES representation is more expressive than SMILES and can represent a broader range of chemical structures. Additionally, SELFIES is more human-readable and can be used to check the correctness of a molecular structure.

SELFIES has been used in various applications such as drug discovery, QSAR modeling, and virtual screening, and it's a powerful tool to handle the complexity and diversity of chemical space. An example of a SELFIES representation of MDMA is \textit{``[C][N][C][Branch1\_1][C][C][C][C][=C][C][=C][C]
[Branch1\_2][Ring2][=C][Ring1][Branch1\_2][O][C][O]
[Ring1][Branch1\_2]''}.

%SELFIES is a 100\% robust molecular string representation. That is, SELFIES cannot produce an invalid molecule, as every combination of symbols in the SELFIES alphabet maps to a chemically valid graph. SELFIES overcomes the drawbacks of all its predecessor in string representation. It does include information about branching, ring, and valence constraints. SELFIES is also an LL(1) formal grammar (or automaton) with derivation rules. The SELFIES grammar is designed with the explicit aim of eliminating syntactically and semantically invalid molecules, for example in generative tasks. A sample SELFIES representation of MDMA is 

\subsection{Graph-Based Representation}

A molecular Graph can be represented as a Matrix for computation purposes. In this representation, each element of the matrix corresponds to a specific atom or bond in the molecule. This matrix representation can be used to encode various types of molecular information, including the topology, stereochemistry, and properties of atoms and bonds. 

One of the main advantages of matrix molecular representation is that it can be easily manipulated and analyzed using mathematical and computational tools. This makes it well-suited for applications such as quantum chemistry calculations, molecular dynamics simulations, and machine learning-based molecular modeling.

A molecule may be represented as a basic molecular network by the graph G = (V, E), where the atoms serve as the nodes, V, and the bonds between them as the edges, e $\in$ E. This encoding can be used to store data on the molecule's architecture. The pairings in E are not ordered because molecular graphs are often undirected. For use in AI applications, molecules are represented by an adjacency matrix, A, where the bond between the nodes $v_{i}$ and $v_{j}$ in the molecular graph G~\cite{Bondy:Graph} is represented by $a_{ij}  \in A$.

%A simple molecular graph of a molecule can be represented as a graph G = (V, E), with atoms as nodes, V, and bonds between each atom as edge e $\in$  E. This representation can be used to encode information about the topology of the molecule. Molecular graphs are generally undirected, meaning that the pairs in E are unordered. For an AI application, molecules are represented by an adjacency matrix, A, where $a_{ij}  \in A$ represents the bond between nodes $v_{i}$ and $v_{j}$  in molecular graph G~\cite{Bondy:Graph}. 

\begin{figure}[htbp]
\centerline{\includegraphics[width=0.9\textwidth, trim={0.5cm 18cm 0.5cm 5cm}, clip]{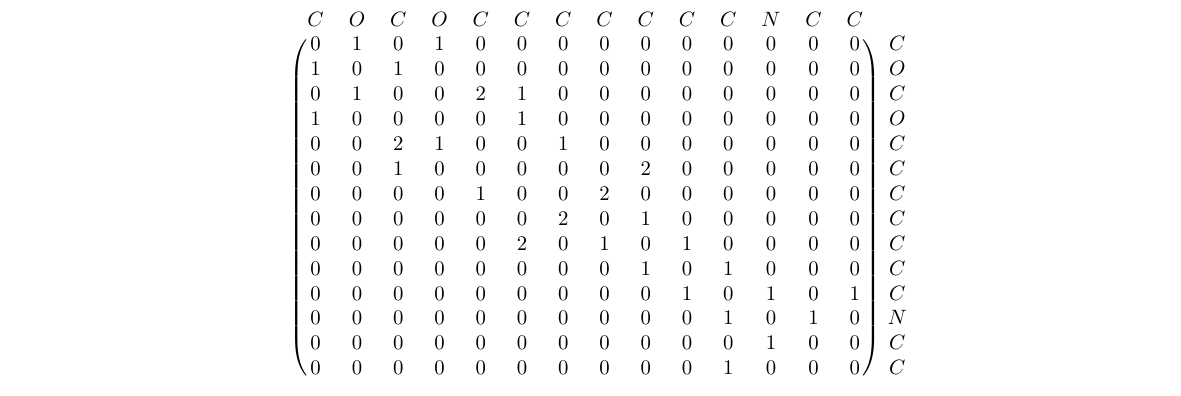}}
\caption{Connectivity Matrix Representation for the MDMA Molecule}
\label{fig:MDMA}
\end{figure}

There are several ways to construct a matrix representation of a molecule, depending on the desired level of detail and the type of information to be encoded~\cite{Michelle:graph}. One of the standard methods is to use a one-hot encoding. For example, in A, if $a_{ij} = 1$ means that there exists a bond between the nodes $v_{i}$ and $v_{j}$ and no bond existence between nodes are represented by $a_{ij}=0$. Another is the distance matrix, where each element represents the distance between two atoms in the molecule. Another approach is to use a connectivity matrix, where each element represents the bond between two atoms. Note that the adjacency matrix does not specify the bond type connecting each pair of nodes. The figure~\ref{fig:MDMA} shows the connectivity matrix representation for the MDMA molecule. Each row and columns represent the node ( atoms in the molecules) of the graph and the vertices of the graph represent the type of bond(connection) between the atoms in the molecules.

% The figure~\ref{fig: acetic} and ~\ref{fig:aceticmat} show the simple molecular graph of acetic acid, and the connectivity matrix representation, whereas the connectivity matrix representation for the MDMA molecule, is shown in figure~\ref{fig:MDMA}. Each row and columns represent the node ( atoms in the molecules) of the graph. in th

% \begin{figure}
% \label{fig:aceticmat}
% \[
% \begin{blockarray}{ccccc}
%    1 & 2 & 3 & 4 \\
% \begin{block}{(cccc)c}
%   0 & 1 & 0 & 0  & 1\\
%   1 & 0 & 2 & 1 & 2 \\
%   0 & 1 & 0 & 0 & 3 \\
%   0 & 1 & 0 & 0 & 4\\
% \end{block}
% \end{blockarray}
% \]
% \caption{Connectivity matrix for Acetic acid}
% \end{figure}

% \begin{figure}[htbp]
% \centerline{\includegraphics[scale=0.3]{acetic_acid.png}}
% \caption{Molecular graph for the Acetic acid}
% \label{fig:acetic}
% \end{figure}

There are several methods to represent the same information. Therefore, the matrix representations covered above are not the only way to represent graphs. For instance, as was already noted, the adjacency matrix's (or atom/bond block's) row order will vary depending on the network traversal technique utilized. Moreover, there is no one right method to represent any molecule when dealing with molecular graphs; the representation used must be suitable for the task.

Matrix representations are not ideal for basic analysis in cheminformatics (e.g., generating a list of compounds, or querying compounds online) because they take up a lot of memory. Recently, however, there has been a move away from SMILES strings to graph representations. It is quite common to add weights to edges, which can be bond lengths, or add parameters to nodes, which can define angles to other nodes, so the graph is a much more flexible representation than a SMILES string. So it is easier to add some information in the Cartesian coordinates without the problems associated with deriving the molecular representation from Cartesian coordinates. 

Moreover, graph representation has demonstrated effectiveness in several other fields where AI techniques like transfer learning are applied. A model that has already been trained for one application can be retrained and used to another task using the AI approach known as transfer learning.

% Matrix representations require a large amount of disk space and are not well adapted for fundamental cheminformatic analysis (i.e., generating a list of compounds, the online query of compounds). However, recently, there has been a shift towards using graph representations over SMILES strings. The reason for the change is  the graph is a much more flexible representation than a SMILES string because it is pretty common to add weights to edges, which could be bond distances, or add parameters to nodes, which could describe angles to other nodes. So, it is easier to add  some information in the Cartesian coordinates without having the problems associated with deriving the molecular representation from Cartesian coordinates.

% Moreover, graph representation has shown successful drive in  many other domains where AI techniques such as transfer learning are used. Transfer learning is  an AI technique where a model pre-trained  for a particular application can be re-trained and used  for another task.

\section{Notable Applications}

Any downstream tasks are carried out by ML models that have been trained using the aforementioned representation as input. Word2Vec served as the inspiration for Mol2Vec, a molecular representation~\cite{Jaeger:Mol2vec}. In Mol2vec, the molecular graph of a molecule is transformed into a sequence of substructures called ``fragments'' and ``bonds''. These fragments and bonds are then treated as ``words'' in a sentence, with the molecular graph serving as the “sentence”. The fragments and bonds are represented as one-hot vectors and passed through a neural network model, which learns to map them to a lower-dimensional space where similar fragments and bonds are closer together in the vector space. The resulting vector representations, known as molecular embeddings, can be used for various machine learning tasks, such as predicting chemical properties or finding similar molecules.

A potential tool for drug development and materials research, Mol2vec has been demonstrated to outperform other state-of-the-art techniques for molecular representation on a number of benchmark datasets. Mol2Vec addresses bit collisions, one of the problems with fingerprints. Smiles2vec learns a representation of a molecule using SMILES tokens using RNN~\cite{Garrett:smiles2vec}. The outputs of the RNN are used to predict molecular characteristics using linear embedding on SMILES characters. Convolutional Neural Networks (CNN) are also useful for predicting chemical characteristics. Yet SMILES string has a drawback in that it can't completely cover the chemical structures and locations of molecules. When a certain structure is not found when training the model, performance suffers.

A particularly interesting approach for automated drug design is using recurrent neural networks (RNNs) as SMILES generators and training them using the learning method known as ``transfer learning''. This includes training the first model on a huge generic data set of molecules to understand the basic syntax of SMILES, then fine-tuning the model on a smaller collection of molecules generated, for example, from a lead optimization program. ~\cite{Amabilino:RNN}. A unique Graph2SMILES model overcomes the SMILE's limitations in depicting molecular structure. The concept is based on the robustness of Transformer models and the permutation invariance of molecular graph encoders~\cite{Tu:Graph2SMILES}.

\section{Summary and Conclusion}

As molecules are complex structures, many different characteristics, including stereochemistry and valence, must be considered in their depiction. The development of bioinformatics and cheminformatics has improved our understanding of molecular activity and made the drug discovery process more effective. Small molecules, polymers, and proteins, as well as the most typical applications for each in AI-driven computational drug discovery, have been covered in this overview.

A molecule can be represented in two different ways: matrix notation and string notation. Both the Matrix and String representations fall short of providing all the details required to characterize a molecular structure. For example, a graph or a SMILES string cannot tell between two isomers that differ only in the orientation of a free O-H group. In conclusion, there are two alternative ways to describe molecules: matrix notation and string notation. Each method has advantages and disadvantages of its own, and depending on the application, one representation may be more appropriate than the other.

\bibliographystyle{unsrt}
\bibliography{srs}

\end{document}